# Seismogenic Potential of the Subduction Zone in Northern Chile

**Sylvain MICHEL[1], Romain JOLIVET[1,2], Jorge JARA[1], Chris ROLLINS[3]**

[1] Laboratoire de Géologie, Département de Géosciences, Ecole Normale Supérieure, PSL Université, CNRS UMR 8538, 24 Rue Lhomond, 75005, Paris, France.
[2] Institut Universitaire de France, 1 rue Descartes, 75006, Paris
[3] GNS Science, Lower Hutt, New Zealand

Sylvain MICHEL : sylvain_michel@live.fr (corresponding author)
Romain JOLIVET : romain.jolivet@ens.fr
Jorge JARA : jara@geologie.ens.fr
Chris ROLLINS: c.rollins@gns.cri.nz

**Key Points:**

- Probabilistic estimate of the seismic potential of the Nazca subduction zone in northern Chile.
- A $M_w$8.8 is the most probable maximum magnitude earthquake considering 4 physical and statistical constraints.
- The potential frictional barrier near Iquique has a limited impact on our estimates considering uncertainties.

The northern Chile region of the Nazca subduction zone ruptured in a ~$M_w$8.5-9.0 earthquake in 1877, which induced a tsunami. The various magnitude estimates of this event are based on the evaluation of historical records, seismic intensities, and/or tide gauge information; however, its actual along-strike extent is debated. Based on geodetic data, previous studies have suggested this region has the potential for a $M_w$8.2-8.8 event. We re-evaluate the seismic potential of the region, accounting for the buildup rate of moment deficit along the megathrust, the earthquake magnitude-frequency distribution, and the physics of earthquakes. We combine an improved probabilistic estimate of moment deficit rate with results from dynamic models of the earthquake cycle, testing in particular the influence of a potential aseismic barrier near the city of Iquique which may influence the extent and magnitude of large events in this region. We find that $M_w$8.8 is the most probable maximum magnitude earthquake in the region and that the potential barrier likely has a limited impact considering uncertainties. We discuss the effect of the moment deficit rate on the fault, the b-value, and the importance of post-seismic processes on our results.

# 1. **Introduction**

Prior to the 1995 Antofagasta earthquake, the last $M > 8$ megathrust earthquake that ruptured the Nazca subduction zone in northern Chile (latitude 25°S to 19°S) occurred in 1877, and the subsequent tsunami devastated the Chilean-Peruvian coast and reached throughout the Pacific (Lomnitz, 2004). The along-strike extent of the earthquake is poorly constrained but most studies agree this event did not rupture south of the Mejillones peninsula (e.g. Kausel, 1986; Métois et al., 2013; Ruiz & Madariaga, 2018). Although debated, the northern extent of this event probably lies between the southern limit of the 1868 $M$~8.8 earthquake and the city of Iquique (Métois et al., 2012; Villegas-Lanza et al., 2016) (Figure 1.a). The 1877 event magnitude has been estimated using historical records (Montessus de Ballore, 1916), seismic intensities (Kausel, 1986; Ramirez, 1988) and tide gauges (Abe, 1979) and ranges from $M$8.5 to 9.0 (Kausel, 1986; Comte and Pardo, 1991; Ruiz and Madariaga, 2018; Vigny and Klein, 2022).

Since then, the region hosted multiple $M$ >7.5 earthquakes (Roth et al., 2017), including four events since 1995 and the advent of geodetic measurements. Those events highlight the potential segmentation of the fault, whether along-strike or along-dip. The 1995 $M_w$8.1 Antofagasta earthquake ruptured almost the full along-dip extent (<50 km depth) south of the Mejillones peninsula (Chlieh et al., 2004; Pritchard et al., 2007), and induced postseismic slip below the co-seismic slip distribution (Figure 1.a) suggesting that no slip deficit is accumulating below ~50 km depth in the area. The 2007 $M_w$7.7 Tocopilla earthquake, located just north of the Mejillones peninsula, ruptured instead between ~30 and 50 km depth, leaving most of the shallow extent of the megathrust locked (Béjar-Pizarro et al., 2010; Schurr et al., 2012). Only a relatively small along-strike portion of the shallow section of the megathrust (~50 km of the event's 150 km length) seems to have hosted post-seismic processes (i.e. aftershocks and afterslip), while the down-dip region from the 2007 co-seismic event stayed relatively quiet (Schurr et al., 2012). Finally, the 2014 $M_w$8.1 Iquique earthquake occurred on the northern part of the megathrust and ruptured two asperities, one shallow and the other at 30-50 km depth (Jara et al., 2018). This earthquake was followed by a $M_w$ 7.7 aftershock 3 days later with a similar along-dip pattern but located south of the co-seismic slip distribution (Duputel et al., 2015).

During the interseismic period, the distribution of locked sections of a megathrust fault, referred to as coupling ($\chi$), can be inferred from geodetic data (e.g. Chlieh et al., 2011; Métois et al., 2012). A coupling of 0 indicates the fault slips at its long-term rate (fully creeping), while a coupling of 1 corresponds to a fully locked fault. Wherever the fault is locked, elastic strain builds up and can be interpreted in terms of a deficit of moment, which can at some point be released through earthquakes or other processes (Reid, 1910; Avouac, 2015a). Evaluating coupling allows one to infer a direct estimate of the moment deficit build up rate (MDR). The latest study of the central Andean coupling distribution based on InSAR (2003-2010) and GNSS data (deployed since 2000) uses a Bayesian approach to probe the probability of whether a point of the fault is locked or not on average over the interseismic period (Jolivet et al., 2020). To first order, the coupled regions of the subduction zone (Figure 1.a) overlap with the co-seismic slip distributions of the $M$ >7.5 earthquakes since 1995, with the exception of the Tocopilla earthquake which potentially ruptured the down-dip creeping-locked transition zone (Schurr et al., 2012). Additionally, at ~20.5°S, Jolivet et al. (2020) infer a reduction in coupling, which may indicate the presence of a frictional barrier (Avouac, 2015a) or a geometrical complexity (Jara et al., 2018). The Iquique mainshock and largest aftershock lie on both sides of this zone of reduced coupling.

The evaluation of the seismogenic potential of a fault includes evaluating the magnitude of the largest potential earthquake (hereafter referred to as the maximum magnitude), $M_{max}$, and its recurrence time, $\tau_{max}$. Considering the uncertainties on the 1877 earthquake along-strike extent and magnitude, the potential maximum magnitude that could occur on the megathrust is poorly constrained. On one hand, according to Chlieh et al. (2011), the moment deficit buildup rate over the approximate region of the 1877 earthquake ([23.5°S,19.0°S] in latitude) is 1.3 x $10^{20}$ N.m/yr. Since 1877, the accumulated moment corresponds to an earthquake of $\sim M_w$8.5-8.8 (Chlieh et al., 2011). On the other hand, Métois et al. (2013) suggest that the 1877 earthquake extended north only up to the city of Iquique (~20.3S) for an area with moment deficit equivalent to only $\sim M_w$8.1-8.3.

However, such seismic potential estimates do not fully consider uncertainties, do not include the micro- to moderate-sized earthquake productivity and do not attempt to consider the physics of earthquakes, especially their potential to propagate along strike. For instance, it has been shown that aseismic regions of the megathrust may control the along-strike extent of large megathrust earthquakes, hence their magnitude (Kaneko et al 2010). In northern Chile, we do not know whether the presence of a low coupling region offshore Iquique significantly impacts the probability of occurrence of large earthquakes. We therefore revisit the seismic potential of the region by using earthquake recurrence models fitting both past earthquake frequencies and moment budget (based here on geodetic data), a strategy that has already been used in previous studies and applied in other regions of the world (Molnar, 1979; Anderson and Luco, 1983; Michel et al., 2018; Mariniere et al., 2021). We do so following a probabilistic approach and we test in particular whether the reduction in coupling observed at ~20.5°S might influence the propagation and thus the magnitude of future large earthquakes. We use statistical results from numerical simulation of the seismic cycle (Kaneko et al., 2010) following a methodology developed, tested and applied in the Himalayas (Michel et al., 2021). This study is thus a direct application of Michel et al. (2021) approach to the Northern Chile subduction zone with an emphasis on the evaluation of the maximum potential earthquake within this region, which is debated, and the role of the potential aseismic barrier offshore Iquique. We provide in addition a sensitivity test of the parameters controlling the earthquake recurrence models to assess their impact on the final results. In the following sections, we first describe the concepts and methods, then describe the data available in northern Chile and apply the methodology to the region before discussing the robustness of our results.

## 2. **Methodology**

The methodology is based on Michel et al. (2021). The period of seismological observation (~100 years) is generally too short to evaluate $M_{max}$ as the recurrence time of large events (M>8) is often centuries to millennia (e.g. Bollinger et al., 2016; Philibosian & Meltzner, 2020). To circumvent this problem, we build seismicity models representing the long-term magnitude frequency-distribution (MFD) of earthquakes, and test them against 4 observational and physical constraint to assess which $M_{max}$ is the most probable under those assumptions. First, the moment release rate from a seismicity model should balance the observed moment deficit rate on the megathrust. Effectively, we evaluate the budget of available moment. Second, the MFD of observed seismicity catalogs should be a possible outcome of the long-term seismicity model, when sampled over the duration of observation (i.e. we assume the MFD is stable over a period longer than the observation period). Third, earthquakes should follow the moment-area scaling law (Kanamori and Brodsky, 2004; Ye et al., 2016). And fourth, earthquakes should be able to propagate through frictional barriers, which we evaluate based on the statistical outcome of simulations of the seismic cycle (Kaneko et al., 2010).

Regarding the seismicity models, we assume they represent MFDs of background seismicity and follow a power-law form up to $M_{max}$ (i.e. Gutenberg-Richter law ; Gutenberg and Richter, 1944, 1954). The MFD models are based upon a non-cumulative power-law MFD truncated at $M_{max}$, which gives rise to a tapered MFD in the cumulative form (i.e. the traditional display when representing the Gutenberg-Richter law; Rollins and Avouac, 2019). Each model is a function of three parameters: (1) $M_{max}$; (2) the recurrence rate of events of a certain magnitude, $\tau_c$; (3) the MFD's *b*-value from the Gutenberg-Richter law (i.e. the relative rate of small and large events), $b$. We also consider a model with a distribution truncated at $M_{max}$ in the cumulative form (i.e. truncated model in Rollins et al., 2019 and Michel et al., 2021) which does not affect our results significantly (Text S1 and Figure S1 and S2).

We assume the studied region is an isolated system. Earthquakes that nucleate within the region cannot propagate out of it and potential earthquakes nucleating outside of it will not propagate within. The results from this study are based on this strong assumption and should be interpreted within this scope.

### 2.1.1. Moment Budget

To evaluate the budget of moment, we compare the moment released by earthquakes with the moment deficit that builds up on the fault during the interseismic period. On one hand, the moment deficit rate, $\dot{m}_0^{def}$, is given as $\dot{m}_0^{def} = \int \mu \, \dot{D}^{def} \, dA$, where $\mu$ is the shear modulus, $\dot{D}^{def}$ is the slip deficit rate, and $A$ is the fault's area. $\dot{D}^{def}$ is linearly related to coupling on the fault, $\chi$, and to the long-term plate rate, $V_{plate}$, as $\dot{D}^{def} = \chi \, V_{plate}$. On the other hand, we estimate the total moment released by earthquakes, $\dot{m}_0^{Total}$, based on the long-term seismicity models. Since the models represent only background seismicity, we add the moment released by aftershocks and aseismic afterslip using a factor $\alpha_s$, which represents the proportion of moment released by background seismicity relative to the total moment released (aftershocks and aseismic afterslip included). The total moment release rate is then $\dot{m}_0^{Total} = \dot{m}_0^{Bckgrd}/\alpha_s$, where $\dot{m}_0^{Bckgrd}$ is the moment rate released by background seismicity. Finally, if $\dot{m}_0^{Total} = \dot{m}_0^{def}$, the model balances the moment budget.

As derived by Rollins and Avouac (2019), and references therein, the cumulative MFD of seismicity models balancing the moment budget directly writes as

$$N(> M_w) = \frac{1 - \frac{2b}{3}}{\frac{2b}{3}} \frac{\alpha_s \, \dot{m}_0^{def}}{m_0^{max}} \left[ \left( \frac{m_0^{max}}{m_0(M_w)} \right)^{\frac{2b}{3}} - 1 \right], \tag{1}$$

where $N(> M_w)$ is the rate of $> M_w$ events, $m_0^{max}$ is the moment released by $M_{max}$, and $m_0(M_w)$ is the moment corresponding to magnitude $M_w$. The probability of a seismicity model balancing the moment budget, $P_{Budget}$, is therefore the combination of probabilities of $M_{max}$, $b$, $\dot{m}_0^{def}$ and $\alpha_s$, for which we sample *a priori* distributions.

### 2.1.2. Observed Magnitude-Frequency Distribution

We evaluate the probability of drawing the observed seismicity catalog (instrumental and/or historical), from the *a priori* distribution of long-term seismicity models over the time period of observation. We assume events are independent, hence we evaluate probabilities assuming that background seismicity follows a Poisson process (Gardner and Knopoff, 1974). For events within a bin of given magnitude $M_i$, the probability to observe $n_{obs}^{M_i}$ events occurring during the time period $t_{obs}^{M_i}$, as characterized by the observed seismicity catalog, assuming the long-term mean recurrence of events is $\tau_{model}^{M_i}$, as defined by a seismicity model, is

$$P_{poisson}^{M_i}\left(n_{obs}^{M_i}, t_{obs}^{M_i}, \tau_{model}^{M_i}\right) = \frac{(t_{obs}^{M_i}/\tau_{model}^{M_i})^{n_{obs}^{M_i}}}{\left(n_{obs}^{M_i}\right)!} e^{-t_{obs}^{M_i}/\tau_{model}^{M_i}}. \tag{2}$$

We then define the probability of the observed seismicity catalog to be an outcome of the long-term seismicity model, $P_{Cat}$, as $P_{Cat} = \prod_i P_{poisson}^{M_i}$.

Effectively, we generate 2500 earthquake declustered catalogs considering magnitude uncertainties and the probability of an event being a background seismicity (e.g. acquired from Marsan et al., 2017, method). The probability $P_{Cat}$ is the mean probability given by the 2500 catalogues.

### 2.1.3. Moment-Area Scaling Law

Global earthquakes statistics show that the moment released by seismic events and the corresponding rupture area are power-law related as $m_0^{seis} \propto A^{3/2}$ (e.g. Kanamori & Brodsky, 2004; Ye et al., 2016). Considering the size of a megathrust fault and the distribution of coupling, all earthquakes are not possible, according to this scaling. We evaluate the probability of occurrence of a given event, $P_{scaling}$, by converting its magnitude to an area using the scaling law, and by checking whether it fits within the seismogenic area. This constraint is applicable for each magnitude examined and does not depend on the seismicity model tested. We account for the rather large uncertainties of the moment-area scaling (see section 3.3).

### 2.1.4. Frictional Barrier Effect

Local reductions in interseismic coupling can be interpreted as frictional heterogeneities acting as barriers to the propagation of earthquakes (Kaneko et al 2010). We evaluate the influence of these potential frictional barriers from the statistics of seismic cycle simulations (Kaneko et al., 2010; Thomas et al., 2014; Michel et al., 2021) based on the rate-state formalism (Dieterich, 1978; Ruina, 1983). Kaneko et al. (2010) relates the probability of a seismic rupture to pass through a frictional barrier to a non-dimensional parameter called the barrier efficiency, $B$. This criterion depends on the fault constitutive parameters and on the dynamics of the rupture:

$$B = \frac{\Delta\sigma_{VS}(a_{VS}-b_{VS})\ln\left(\frac{V_{Dyn}}{V_i}\right)D_{VS}}{\beta\Delta\tau_{VW}D_{VW}}, \qquad (3)$$

where $a_{VS}$ and $b_{VS}$ are constitutive parameters of the rate-and-state friction law at the barrier, and $\Delta\sigma_{VS}$ and $D_{VS}$ are the barrier's effective normal stress and width, respectively. $V_i$ corresponds to the interseismic slip rate, and $V_{Dyn}$ to the rupture's slip rate. $D_{VW}$ and $\Delta\tau_{VW}$

### 2.1.5. Seismicity model Probability, $P_{SM}$

We define the probability of a seismicity model, $P_{SM}$, as the product $P_{SM} = P_{Budget}\, P_{Cat}\, P_{scaling}\, P_{Barriers}$ (Michel et al., 2021) calculated from the imposed constraints (section 2.1). Based on $P_{SM}$, we then estimate the marginal probabilities of $M_{max}$ and of the *b*-value, respectively $P_{M_{max}}$ and $P_b$. Defining $M_{mode}$ as the magnitude at which $P_{M_{max}}$ peaks, we evaluate the probability $P(\tau\,|\,M_w = M_{mode})$ of the rate of $M_w = M_{mode}$ events, which accounts for all earthquakes from all of the models (i.e. not only $M_{max}$).

## 3. Data and Uncertainties

In order to apply this methodology and evaluate the seismicity models, we explore and sample the constraint parameters within their uncertainties. In this section, we present the data used in this study, and describe how uncertainties are estimated.

### 3.1. Seismicity Model Constitutive Parameters, Moment Deficit Rate and Seismogenic Zone

Due to Eq. (1), we can limit the seismicity model parameter space (section 2.1) to be explored, examining the MDR instead of $\tau_c$. The parameter space $M_{max}$, $b$, $\alpha_s$ and MDR is then explored through a grid search. $M_{max}$ and $b$ are sampled uniformly over $M_{max} \in \mathcal{U}(7.8,10)$ and $b \in \mathcal{U}(0.05,1.50)$, respectively. The PDF of $\alpha_s$ is assumed Gaussian with $\mathcal{N}(80\%, 20\%)$ based on global statistics of earthquake post-seismic behavior (Alwahedi & Hawthorne, 2019; Avouac, 2015b).

We use the coupling model from Jolivet et al. (2020) to calculate the MDR ($\dot{m}_0^{def}$) and its related uncertainty (see section 2.1.1). Jolivet et al. (2020) have sampled the PDF of the coupling models that fit the geodetic data, hence providing 245760 coupling models based on GNSS and InSAR data. From these models, we build the PDF of moment deficit rate $\dot{m}_0^{def}$ following the methodology from section 2.1.1. The PDF follows a Gaussian distribution with $\mathcal{N}(1.27\ 10^{20}, 5.48\ 10^{18})$ N.m.yr$^{-1}$.

The seismicity model parameter space grid search is proceeded using steps of 6.36 10$^{17}$ N.m.yr$^{-1}$ for the MDR, 0.1 of magnitude for $M_{max}$, 0.01 for the $b$-value, and 10% for $\alpha_s$.

Additionally, we define the seismogenic zone along strike extent and width within an iso-coupling $\chi_{thresh}$ of 0.3. We then discretize the model in thirty seven 17.5 km-long segments and compute the average and standard deviation coupling and width of the seismogenic zone in each bin (Figure 1).

### 3.2. Seismicity Catalog

We combine two earthquake catalogs, an instrumental one from the *International Seismological Center* (ISC; Willemann and Storchak, 2001; Bondár and Storchak, 2011; Storchak et al., 2017, 2020; International Seismological Centre, 2022), and an aggregate one from Roth et al. (2017) including both instrumental and historical events.

The instrumental ISC catalog is itself a combination of catalogs. Most of the seismic events in northern Chile originate from the *Centro Sismológico Nacional* catalog (Barrientos, 2018). The type of magnitude in the ISC catalog is heterogeneous. We thus convert the magnitude of each event into moment magnitude ($M_w$; see Text S2 and Figure S3 and S4). We then build two separate catalogs, associated with two different periods of observation, that we decluster using the method of Marsan et al. (2017), following the parametrization described by Jara et al. (2017) for the region. For the first catalog, we consider events between 2001 and 2021, with an estimated magnitude of completeness ($M_c$) of 3.3 (See Text S2). From the resulting declustering of the catalog, we select events from 2010 (date from which background seismicity rate appears constant), and within 200 km east from the trench to avoid earthquakes related to the orogenic processes within the Andes at ~100 km depth (Figure S5). For the second catalog, we take instead events between 1970 and 2021, with an estimated $M_c$ of 5.7. After declustering, we select events from 1980 and also within 200 km east from the trench (Figure S6). For both catalogs, we do not discriminate between events attributed to slip on the subduction interface and events rupturing structures within the overlying forearc. The elastic loading resulting from coupling along the interface induces strain over a large area and we assume that seismicity in the arc only results from the corresponding stresses.

The other earthquake catalog is an aggregate of various catalogs combined by Roth et al. (2017). $M > 6.5$ earthquakes between 1500 and 1899 are from the CERESIS catalog (Giesecke et al., 2004), events between 1900 and 2009 from ISC-GEM data (Storchak et al., 2013), and earthquakes after 2009 (which consist of a $M6.5$ in 2009 and of the $M_w 8.1$ Iquique earthquake) are included manually (Duputel et al., 2015; Jara et al., 2018). We do not include the largest aftershock of the Iquique earthquake as we use a declustered catalog. We select events since 1877 within the region, excluding the 1877 event, and assume that events spatially straddling previous earthquakes within a 1.5 yr period are aftershocks, and are thus removed. The events magnitude uncertainties are fixed to 0.2 for events before 1995, and 0.1 thereafter. Roth et al. (2017) suggest a $M_c$ of about 7.0 while a maximum curvature method (Wiemer and Wyss, 2000) evaluates this magnitude of completeness around 7.25. Note that when applying the observational seismicity catalog constraint (section 2.1.2), we only compare events with $M > 4.25$, $M > 6.25$ and $M > 7.75$ for the first and second catalog from ISC, and the aggregate catalog from Roth et al. (2017), respectively, to avoid border effect from the magnitude of completeness when exploring magnitude uncertainties (Felzer, 2008).

### 3.3. Additional Parameters needed to Implement the Scaling Law and Frictional Barrier Constraints

Constraints from the earthquake scaling law and frictional barrier are explored separately from the other two constraints by testing 500 000 events sampled uniformly between $M6$ and $M10$. Considering the scaling law, we use the moment-area scaling and associated uncertainty modeled by Michel et al. (2021) derived from the database of large subduction earthquakes of Ye et al. (2016). We evaluate a linear relationship between the rupture area, $A$, and seismic moment of earthquakes, $m_0$, such as $Log_{10}(m_0) = \frac{3}{2} A + Q$. Least squares regression suggests $Q = 15.15$ (with $A$ in km²). Earthquake moment is normally distributed around this trend with a standard deviation of $\sigma_{scaling} = 0.23$ (Michel et al., 2021), used as the uncertainty on the scaling law in our exploration. Other studies yield different uncertainty estimates on the scaling. Using data from Leonard (2010), $\sigma_{scaling}$ reaches 0.5, roughly twice than the one estimated on data from Ye et al. (2016). Such larger uncertainties directly impacts $P_{scaling}$ and thus $P_{M_{max}}$ (Figure S7).

To apply the frictional barrier constraint in a probabilistic manner, we need to evaluate the values and uncertainties of the constitutive, dynamic and geometric parameters involved in the barrier efficiency. To our knowledge, there are no estimates available for the constitutive parameters $\Delta\sigma_{VS}(a_{VS} - b_{VS})$ in the northern Chile region of the Andean subduction zone. We assume values and uncertainties from other parts of the subduction zone, including a post-seismic study from Weiss et al (2019) and Franck et al (2017). The interseismic slip rate, $V_i$, is taken from the coupling model. The rupture properties such as seismic slip velocity, $V_{Dyn}$, and stress drop, $\Delta\tau_{VW}$, are taken from the literature (Kanamori & Brodsky, 2004, Cocco et al., 2016). The length of an event, $D_{VW}$, depends on the procedure described by Michel et al., 2021. All parameters are summarized in Table S1.

We assume that there is only one potential barrier located offshore Iquique. The barrier width, $D_{VS}$, and related uncertainty are based on the along-strike distribution of coupling and the coupling threshold $\chi_{thresh}$ (see section 3.1). From the PDF of coupling, the distribution of possible widths of the barrier is a positive-truncated Gaussian $\mathcal{N}(5.3, 21.9)$ km, with a non-negligible probability of being inexistent (Figure S8).

Finally, the relationship between the barrier efficiency ($B$) and the probability that an event passes a barrier is itself prone to uncertainty. This relationship is indeed based on the statistical outcome from 2D numerical models of the seismic cycle (Kaneko et al., 2010). We model the relationship with a linear regression for $B \in [0\ 1.2]$, and impose a probability equal to 0 for $B > 1.2$ (Figure S9). The distribution of the probabilities relative to the model follows to a Gaussian with a standard deviation of 9.3 % (Michel et al., 2021; Figure S9), that we will use as uncertainty for the relationship.

## 4. Results

The probability of a given seismicity model, $P_{SM}$, is shown in Figure 2, with the corresponding marginal probability of the maximum magnitude of an earthquake $P_{M_{max}}$. Maximum magnitude is approximately normally distributed and peaks at $M_w 8.85$ (median and mean of $P_{M_{max}}$ can be found in Table S2).

Earthquakes larger than $M_w$ 9.25 are very improbable (less than 1% considering $P_{M_{max}}$; Figure 2). The sharp decrease of $P_{M_{max}}$ for magnitudes larger than 9.2 is mainly controlled by the moment-area scaling law (Figure 3 and 4.a), and to a lesser extent by the barrier effect (section 2.1.4). The barrier effects only affects magnitudes larger than $M_w$ 8, cutting the probability of an event by ~50% above magnitude 9 (black line in Figure 3.a). The effect of the other parameters of the barrier efficiency are shown in Figure S10. $P_b$, the marginal probability of the *b*-value, is also normally distributed, peaks at ~0.8 and reaches a probability close to 0 at $b$ =0.9 (inset in Figure 2). $P(\tau \mid M_w = M_{mode})$, the probability of the rate of $M_w = 8.85$ events, implies a recurrence time of such events of between ~1000 and ~6300 yrs, with a peak at ~2500 yrs (Figure 2 and 4.b).

The probability of occurrence of an event with $M > M_w$ for a time period $T$, $P(M > M_w \mid T)$, is an estimate frequently used in seismic hazard analysis. For northern Chile, the probability of having at least a $M_w > 8.8$ in a 30, 100, 1000 and 10 000 yr period is of ~1, ~4, ~29 and ~63% (Figure 4.c). One could expect a probability closer to 1 for $T = 10\,000$ yr considering the recurrence time at $M_{mode}$, however, a bit less than half of the probable seismicity models have a $M_{max}$ below 8.7 and have thus a probability equal to 0 for producing $M_w > 8.7$ events.

Another estimate of importance for seismic hazard analysis is the probability of the distance between a chosen site and the seismic source. We evaluate this probability, taking the barrier effect into consideration, using the procedure from section 2.1.3 and 2.1.4, as we are able to calculate the probability of a point of the fault to be part of an event of magnitude $M_w$. Figure 4.d shows the difference between the PDFs of $M_w > 8.5$ events location with and without the barrier effect. The northern segment of the subduction zone, past the city of Iquique, is less prone to be part of a $M_w > 8.5$ due to the barrier. We can then, for example, calculate the probability of the distance between the cities of Iquique or Tocopilla, and a $M_w > 8.5$ seismic source, taking into account the potential frictional barrier (Figure 4.e). In this study we evaluate only the distances relative to one seismic source (i.e. the subduction interface), as opposed to the common use in PSHA that takes into consideration all sources. Nevertheless, the results from this study's approach could eventually be integrated into such seismic hazard analysis.

## 5. Discussion

### 5.1. Influence of the parameters of the long-term seismicity model

Under all the assumptions supporting our analysis, our main result suggests that the most probable maximum magnitude earthquake is 8.8 and that earthquakes larger than Mw 9.1 are very improbable in the region (less than 2% considering $P_{M_{max}}$) assuming that the region is an isolated system (i.e. earthquakes cannot propagate out of it and external events cannot propagate within it). In addition, the introduction of a barrier to the propagation of earthquakes and of the scaling law leads to a lowering of the probability of an earthquake rupturing the entire northern Chile seismic gap (from the Mejillones peninsula to the Arica bend). It appears that the barrier effect is less influential than the earthquake scaling law due to the large uncertainties of the barrier efficiency parameters (Section 2.1.4 and 3.3), and the fact that the barrier is located on one side of the fault and only separates a small portion of the fault. Before introducing these physics-based constraints, the probability of a seismicity model ($P_{SM}$), and thus $P_{M_{max}}$, depends on the MDR, *b*-value and $\alpha_s$. We explore the influence of each of these parameters on $P_{M_{max}}$ without accounting for the scaling law and barrier constraints in order to evaluate the effect of the parameters at the core of our method. We will mostly refer to the mode of $P_{M_{max}}$ but the median and mean of the PDF for each of the following scenarios can be found in Table S3.

First, we explore the effect of the estimate of the moment deficit rate. We compare three case: (1) a megathrust fully locked down to 60 km depth, (2) the MDR from Jolivet et al. (2020) divided by 2 (keeping the same uncertainty), and (3) the MDR of Jolivet et al. (2020) divided by 5 (Figure 5.a). Those scenarios are quite radically different from our setup and have a strong effect on $P_{M_{max}}$. While the fully locked fault results in a PDF peaking at $> M_w 10$, lower moment deficit rate tends to saturate at $M_w 8.3$, constrained by the aggregate earthquake catalog (seismicity models with $M_{max}$ below the observed maximum magnitude of the catalog will be highly improbable).

Second, we explore the influence of the assumed moment contribution of post-seismic and aseismic deformation. Figure 5.b shows tests for $\alpha_s$=0.8, 0.67, 0.50 and 0.40, which corresponds to a ratio between postseismic processes and background seismicity moment release of 0.25, 0.5, 1.0 and 1.5, respectively (Note that in our main model, $\alpha_s$ is normally distributed around a mean of 80% with a standard deviation of 20%). Decreasing $\alpha_s$, which increases the percentage of moment released by post-seismic processes relative to background seismicity, decreases the amount of moment released by background seismicity needed to balance the moment budget, thus $M_{max}$.

Third, we explore the effect of the *b*-value (Figure 5.c). Fixing this parameter to a single value is a strong assumption due to the observed seismicity catalog constraint, and will have an important impact on $P_{M_{max}}$. The lower the *b*-value, the smaller the probable $M_{max}$. However, seismicity models with *b*-values around 0.8 will fit more easily the observed MFD of earthquakes (most probable *b*-value using only the moment budget and catalog constraints).

In conclusion, the MDR is relatively well constrained observationally from geodetic data. Although end-member hypothesis show significant influence on the results, we can confidently say that such hypothesis are quite improbable. The *b*-value has also a strong influence but is constrained by the earthquake catalogs. The large uncertainties on the magnitude of the catalog events (up to 0.5) allows for a broad possible range of *b*-value, between 0.7 to 1.0 before applying the scaling law and barrier constraint, and between 0.7 and 0.9 after. For the poorly constrained parameter, namely the ratio of post- versus co-seismic slip, we observe that significant variations in $\alpha_s$ changes the most probable maximum magnitude to some extent with variations contained between 8.95 and 9.55. However, adding the scaling law constraint diminishes the impact of $\alpha_s$ restraining instead $P_{M_{max}}$ mode between 8.75 and 8.95 thus not affecting our conclusions. Note that the two types of long term seismicity models explored here (i.e tapered and truncated models; see Section 2, Text S1, and Figure S1 and S2) provide similar results. This is not necessarily the case in other regional setting (e.g. in the Ecuador-Colombian subduction; Mariniere et al., 2021).

We finally explore the influence of the coupling threshold within which we define the seismogenic zone. This threshold has a small impact on the moment-area scaling law constraint, shifting $P_{scaling}$ of 0.05 magnitude when taking coupling thresholds of 0.2 or 0.4 (red lines in Figure 3.a). An extreme scenario which we already explored above would be to assume the fault completely locked (coupling equal to 1) down to 60 km depth. Such scenario would significantly increase the width and along-strike extent of the seismogenic zone. Such scenario results in $P_{M_{max}}$ starting to increase at $M_w$8.75, peaking at $M_w$9.15, and dying out at $M_w$9.45 (Figure S11), thus larger $M_{max}$ than when using the coupling map. However, such scenario is not consistent with geodetic data over the interseismic period (Métois et al 2016, Jolivet al 2020).

## 5.2. Comparison with earlier estimates of seismogenic potential

Chlieh et al. (2011) estimate a moment deficit rate of 1.3 $10^{20}$ N.m/yr over their proposed 1877 earthquake spatial extent, roughly from the city of Arica in the North to the city of Antofagasta in the south. Their estimate slightly overestimates ours using roughly the same area (9.1 $10^{19}$ ± 0.4 $10^{19}$ N.m/yr) for a difference in equivalent magnitudes of ~0.1. Assuming that the 1877 earthquake extended only up to the city of Iquique, Métois et al 2013 evaluate moment deficit rate of ~1.5-2.9 $10^{19}$ N.m/yr, corresponding to an equivalent magnitude of $M_w$8.1-8.3. Within this area, the MDR from the coupling map from Jolivet et al. (2020) indicates ~6.2 $10^{19}$ N.m/yr, which will produce a difference of 0.3 in terms of magnitude.

The magnitude of the 1877 event was actually recently re-estimated as an $M_w 8.5$ based on the reevaluation of intensity maps and tsunami heights by Vigny and Klein (2022). This value is on the edge of the PDF of $P_{M_{max}}$ we estimate (see section 4), and the recurrence time from Vigny and Klein (2022) differs from ours (~150 yrs instead of ~800 yrs for $M_w 8.5$ events). Uncertainties, whether on the magnitude or recurrence time, are in any case relatively large (Figure 2).

### 5.3. Megathrust segmentation

We interpreted the low coupling near the city of Iquique (Figure 1.a) as a frictional barrier, assuming a local variation in rheological properties of the megathrust are responsible for this reduction in coupling. However, other complexities might be at the origin of such feature (Jara et al., 2018; Maksymowicz et al., 2018). For instance, geometry or fluids are known to influence the dynamics of slip along faults (Romanet et al 2018; Noda & Lapusta, 2010; Sibson, 1973). One would therefore need to reconsider the probability of an earthquake rupturing such low coupling region under the physical constraints imposed by these mechanisms.

We also note that the 1877 event probably stopped south at about the northern extent of the 1995 Antofagasta earthquake, while no drop in coupling is observed at this location (Figure 1.a). This apparent segmentation might originate from the stress shadow imposed by an earthquake that ruptured the same region than the 1995 Antofagasta earthquake, although this is speculative. In any case, this highlights the role of historical seismicity and, in particular, the slip distribution of past earthquakes, which we do not account for in this methodology. Indeed, background seismicity, including $M_w$>7.5 earthquakes, are viewed as independent events. Such assumption is probably not applicable for large events as their occurrence is surely dependent on their interactions in time and space. Finally, our study only considers along-strike heterogeneity while there is evidence for along-dip segmentation (i.e. Jara et al., 2018; Lay et al., 2012; Ruiz & Madariaga, 2018).

Finally, we acknowledge that our study only accounts for a finite region of the subduction zone. Large earthquakes could effectively propagate out of this region as it has been recently proposed based on geo-archeological evidence which suggests the potential occurrence of a M9.5 event 3800 yrs ago (Salazar et al., 2022).

# 6. Conclusion

We propose a probabilistic evaluation of the seismogenic potential of the northern-Chile subduction zone taking into account the moment budget of the megathrust, the MFD of observed earthquakes, the moment-area scaling law, and the effect of a potential frictional barrier. We find that, given the uncertainties, a $\sim M_w 8.8$ event is the most probable maximum magnitude earthquake (between ~8.5 and 9.1 considering the 5 and 95 percentile of the cumulative PDF of $P_{M_{max}}$), that events with such magnitude tend to occur every ~2500 yrs if taking all probable seismicity models into account (between 1500 and 6300 yrs considering the 5 and 95 percentile of $P(\tau \mid M_w = 8.85)$), and that the potential barrier near the city of Iquique has a limited impact considering its uncertainty. The methodology presented in this study does not take into account the history of large events, and is thus time independent. Since the 1877 earthquake, the moment deficit accumulated on the fault north of the city of Antofagasta is equivalent to a $M_w 8.6$. However, the 2014 Iquique earthquake and its aftershock might have released moment deficit at its location and potentially left a stress shadow that might hinder the propagation of future large events. Nevertheless, following this scenario, the moment deficit accumulated between the city of Antofagasta and Iquique since 1877 is equivalent to a $M_w 8.5$.

## Data and Resources

We use the instrumental seismicity catalogue from the International Seismological Center (http://www.isc.ac.uk/iscbulletin/search/catalogue/). The historical seismicity data is available through Roth et al., 2017. The coupling model data is available through Jolivet et al., 2020. The moment and area data of subduction earthquakes is taken from Ye et al., 2016. The relationship between the barrier efficiency and the probability of an earthquake to pass a frictional barrier is given by Kaneko et al., 2010. The supplementary material contains (1) a description of the results of northern Chile seismogenic potential using a truncated seismicity model (Text S1), (2) information about the process followed to decluster the instrumental catalog (Text S2), (3) figures illustrating the difference between the tapered and truncated seismicity models (Figure S1 and S2), (4) figures describing the magnitude type conversion of the instrumental catalog events into moment magnitude (Figure S3 and S4) (5) figures describing the instrumental catalog and its declustering (Figure S5 and S6), (6) figure showing a comparison between using Leonard (2010) and Ye et al. (2016) data on the scaling law constraint (Figure S7), (7) figures describing the uncertainties on the size of the potential barrier near the city of Iquique and on the probability of an event to pass a barrier (Figure S8 and S9), (8) a figure showing the effect of the frictional barrier constraint on the parameters of the barrier efficiency (Figure S10), (9) a figure showing the results of northern Chile seismogenic potential taking a fault fully coupled down to 60 km depth (Figure S11).

## Acknowledgements and Data

This project has received funding from the European Research Council (ERC) under the European Union's Horizon 2020 research and innovation program (project Geo-4D, grant agreement 758210). RJ acknowledges funding from the Institut Universitaire de France. The study complies with FAIR Data standards. We thank the Centro Sismológico Nacional de Chile (CSN, www.csn.uchile.cl) and its researchers and technicians for producing, maintaining, and providing the earthquake catalog used in this study.

## Declaration of Competing Interests

The authors acknowledge there are no conflicts of interest recorded.

This is the author accepted version of the manuscript. For the published version: https://doi.org/10.1785/0120220142

Ye, L., T. Lay, H. Kanamori, and L. Rivera, 2016, Rupture characteristics of major and great ( $M_w \geq 7.0$) megathrust earthquakes from 1990 to 2015: 2. Depth dependence, J. Geophys. Res. Solid Earth, 121, no. 2, 845–863, doi: 10.1002/2015JB012427.

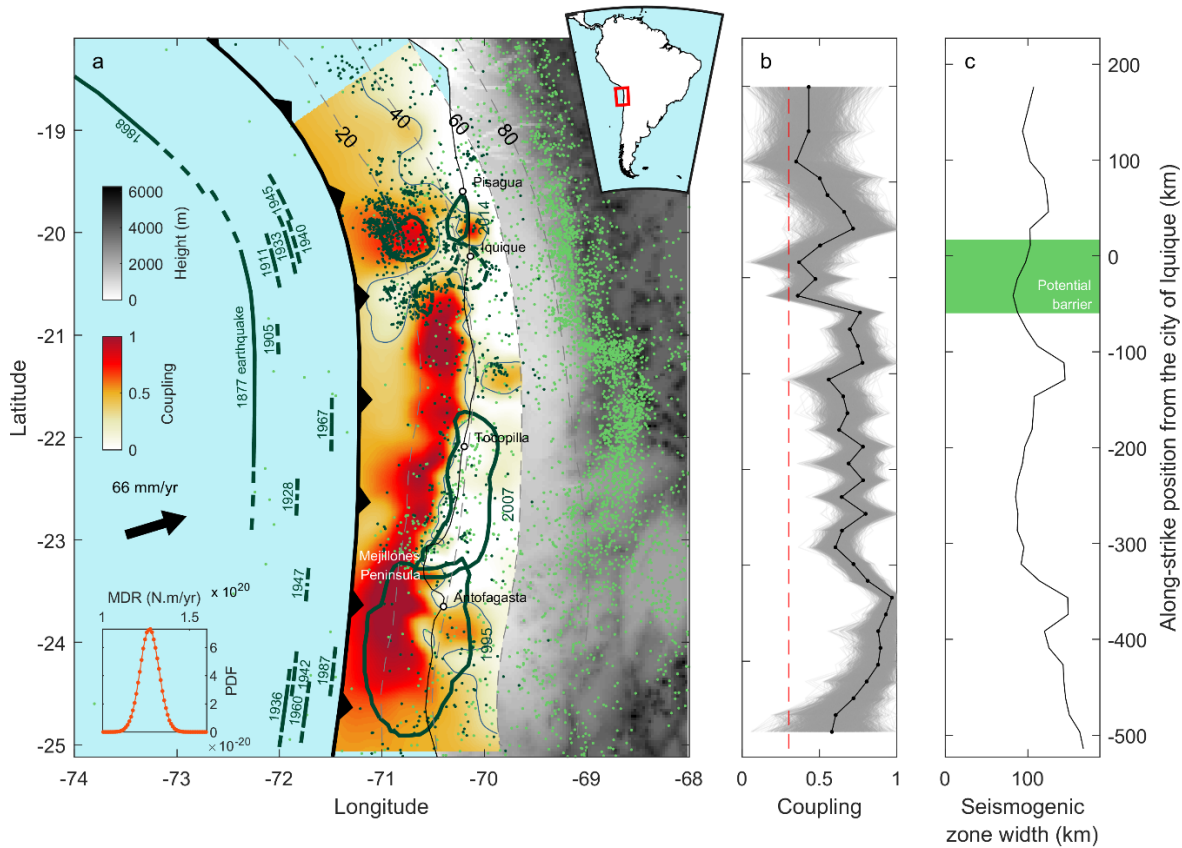

Figure 1: Regional setting of the northern Chile region of the Nazca subduction zone. (a) Interseismic coupling map (Jolivet et al., 2020). Dots indicate microseismicity since 1995 from the CNS catalog. Dark green dots are earthquakes selected for the seismic potential analysis (<200 km from the subduction trench). Thick green lines offshore indicate the spatial extent of historical large earthquakes ($M_w > 7$ between 1877 and 1995; Roth et al., 2017). Co-seismic slip distribution of recent large earthquakes (since 1995) are indicated by the green contours. The thin solid blue line on the coupling map delimits the bottom extent of the coupled zone, based on a coupling threshold of 0.3. Iso-depths of the subduction interface are indicated by the thin gray dashed lines (in km). The bottom left inset shows the PDF of the moment deficit rate. (b) Along-strike distribution of coupling averaged along-dip of each of the 245 760 coupling models (gray lines). The black line is the mean value of the 245 760 coupling models. The coupling threshold of 0.3 used in this study is indicated by the red dashed line. (c) Width of coupled zone based on a coupling threshold of 0.3. The potential location of the aseismic barrier is indicated by the green shading.

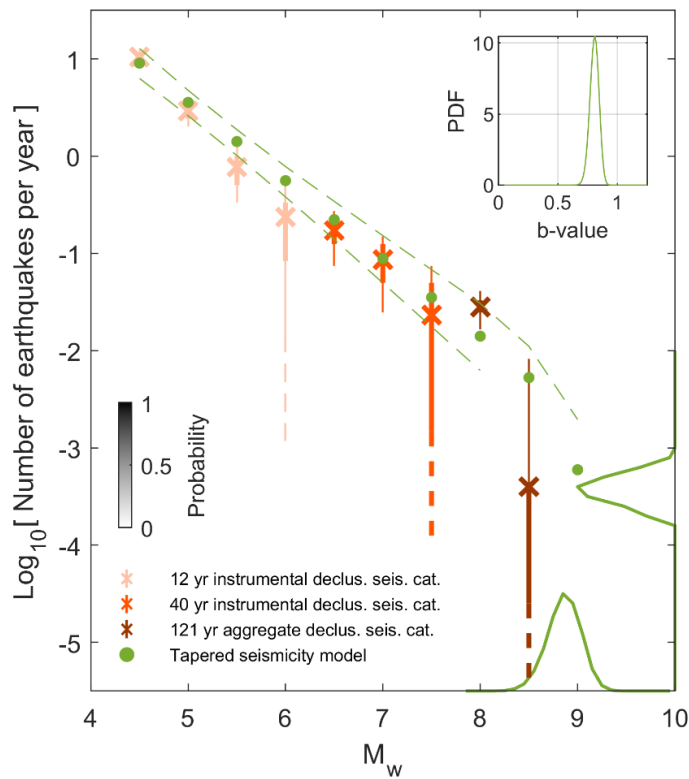

Figure 2: (a) Seismic potential analysis using all constraints: moment budget, observed magnitude-frequency distribution, moment area scaling law, and the frictional barrier effect. The rate of occurrence of earthquakes, within their observation period, are indicated by brown (Roth et al., 2017, catalog; events from 1850 to 2022), orange (*International Seismological Center* (ISC) catalog; 1980-2022 events) and pink (ISC catalog; 2010-2022 events) crosses. The associated thin and thick vertical lines correspond to the catalogs 15.9-84.1% (1-sigma) and 2.3-97.7% (2-sigma) quantiles, respectively. Green dots show the mean of the marginal PDF of the long-term seismicity models. Green dashed lines indicate the spread of the 1% best seismicity models. The marginal probability of $M_{max}$, $P_{M_{max}}$, is indicated by the solid green line on the $M_w$ axis. The solid green line on the earthquake frequency axis indicates the probability of the rate of events, $\tau$, with magnitude $M_w = 8.85$, thus $P(\tau \mid M_w = M_{Mode})$, which considers all magnitudes in the seismicity models and not only the recurrence rate of $M_{max}$. The top-right inset shows the marginal probability of the b-value. Note that the seismicity MFDs in the figure are not in the cumulative form.

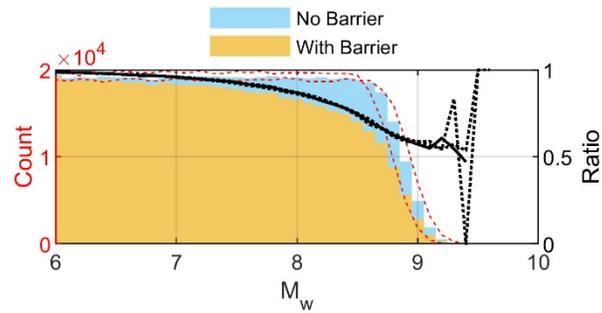

Figure 3: PDFs of $M_w$, expressed in counts, considering only the moment-area scaling law (blue histogram) and considering both the scaling law and potential frictional barrier into account (yellow histogram) The PDFs are evaluated from 880 000 events uniformly sampled between $M_w$ 6 and 10. The effect of the potential frictional barrier, alone, is estimated taking the ratio between the yellow and blue histograms (solid black line). The dashed black lines represent the effect of the potential frictional barrier for a coupling threshold equal to 0.2 and 0.4 (instead of 0.3). The red dashed lines indicate the position of the PDF, using the moment-area scaling law constraint only, if taking a coupling threshold equal to 0.2 and 0.4.

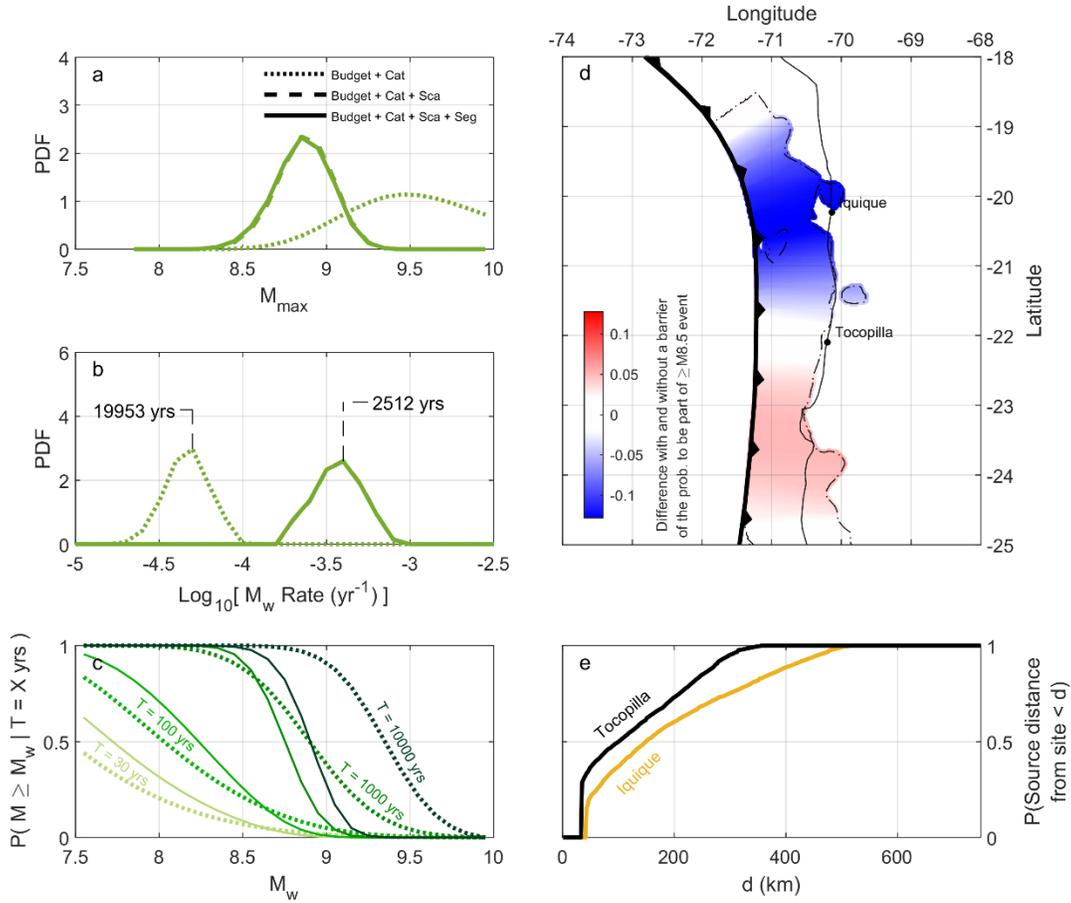

Figure 4: (a) Marginal PDF of $M_{max}$ for three combinations of constraints. (b) Same as (a) but for the marginal PDF of the recurrence time of events with $M_w = M_{Mode}$, $P(\tau \mid M_w = M_{Mode})$. (c) Probability of occurrence of earthquakes of magnitude larger than $M_w$ over a period of X yrs. We show the probability of occurrence of such events for 4 time periods, 30, 100, 1e3 and 1e4 yrs. In (a), (b) and (c), dotted lines represent the marginal PDFs considering both the moment budget and seismicity catalog constraint, the dashed lines indicate the PDFs when adding the earthquake scaling constraint (not shown in (c)), while the continuous lines indicate the PDFs using all constraints. (d) Difference of the probability of rupture extent with and without the barrier for $> M_w 8.5$ earthquakes using only the scaling law and the potential frictional barrier constraints into account. (e) Cumulative PDF of $> M_w 8.5$ source distance from the cities of Iquique and Tocopilla.

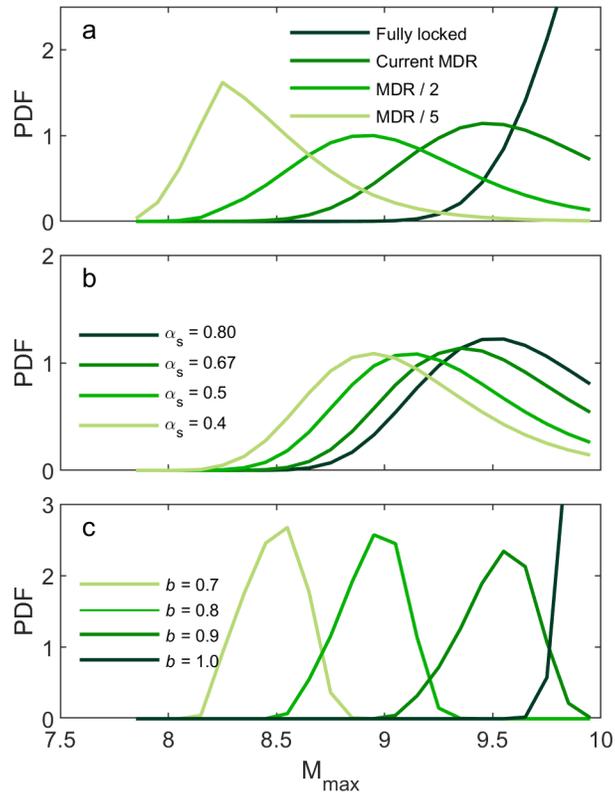

Figure 5: Marginal PDF of $M_{max}$, $P_{M_{max}}$ for different values of (a) moment deficit rate, (b) $\alpha_s$, the proportion of moment released by aftershocks and aseismic afterslip relative to the total moment released, (c) *b*-value.